# Majorization: Here, There and Everywhere

**Barry C. Arnold**

## Dedication

This article is written for Ingram Olkin on the occasion of his 80th birthday. Ingram has provided inspiration for me over the last 40 years and continues to inspire. I am indebted to him for his encouragement and support throughout my career. I am contributing this humbly in the sure knowledge that he could have written it better than I.

*Abstract.* The appearance of Marshall and Olkin's 1979 book on inequalities with special emphasis on majorization generated a surge of interest in potential applications of majorization and Schur convexity in a broad spectrum of fields. After 25 years this continues to be the case. The present article presents a sampling of the diverse areas in which majorization has been found to be useful in the past 25 years.

## 1. INTRODUCTION

Prior to the appearance of the celebrated volume *Inequalities*: *Theory of Majorization and Its Applications* (Marshall and Olkin, 1979) many researchers were unaware of the rich body of literature related to majorization that was scattered in journals in a wide variety of fields. Indeed, many majorization concepts had been reinvented and often rechristened in different research areas (e.g., as Lorenz or dominance ordering in economics), complicating the difficulties for the researcher when trying to relate current research to the extant corpus. Of course, the appearance of the Marshall and Olkin volume changed all that. They heroically had sifted the literature and endeavored to arrange ideas in order, often providing references to multiple proofs and multiple viewpoints on key results, with reference to a variety of applied fields. Many of the key ideas relating to majorization were already discussed in the (also justly celebrated) volume entitled *Inequalities* by Hardy, Littlewood and Pólya (1934). Indeed, this slim volume still merits occasional revisits since there remain in it many "seedlings for further research" (to borrow Kingman's apt descriptive phase). Of course the Hardy, Littlewood and Pólya volume, though slim and printed on small pages, was all meat and no gravy: more like a series of insightful telegrams. Only a relatively small number of researchers were inspired by it to work on questions relating to majorization.

But things were different after 1979. Marshall and Olkin sold the product much more effectively. Whenever a situation was encountered in which a solution or an extreme case involved a discrete uniform distribution, the possibility of a majorization proof was now apparent if not to all, certainly to many, and certainly in many different areas of research. Moreover, if a uniform allocation or distribution was in a sense optimal, then the concept of majorization frequently could be used to order competing allocations or distributions.

*Barry C. Arnold is Professor, Department of Statistics, University of California, Riverside, California 92521, USA e-mail: barry.arnold@ucr.edu.*

Naturally extensions of the majorization concept were possible and indeed many have been fruitfully introduced. The focus of the present article is, however, on classical majorization. The goal is to provide a hint (via selected examples from the post-1979 literature) of the vast array of settings in which majorization provides a useful and interpretable ordering. In no sense can such a survey be complete. I apologize, in advance, to researchers who, quite legitimately, can point to papers of their own which they feel would be even better illustrations of the theme: Majorization, here, there and everywhere. Nevertheless it is my hope that the examples selected will be found to be interesting, to be sufficiently diverse in order to illustrate the potential ubiquity of dispersion ordering (a.k.a. majorization) concepts and, perhaps, to inspire researchers to seek even more research niches in which majorization and Schur convexity will play a useful role.

## 2. SOME NEEDED DEFINITIONS

We will say that a vector $\underline{x} \in \mathbf{R}^n$ majorizes another vector $\underline{y} \in \mathbf{R}^n$ and write $\underline{x} \succ \underline{y}$ if for each $k = 1, 2, \ldots, n-1$ we have

$$\sum_{i=1}^{k} x_{i:n} \leq \sum_{i=1}^{k} y_{i:n}$$

and

$$\sum_{i=1}^{n} x_{i:n} = \sum_{i=1}^{n} y_{i:n}.$$

In the above we denote the ordered coordinates of a vector $\underline{x} \in \mathbf{R}^n$ by $x_{1:n} \leq x_{2:n} \leq \cdots \leq x_{n:n}$.

A function $g : \mathbf{R}^n \to \mathbf{R}$ is said to be Schur convex if $\underline{x} \succ \underline{y}$ implies $g(\underline{x}) \geq g(\underline{y})$. For additional details and alternative characterizations of majorization and Schur convexity, we naturally refer to Marshall and Olkin (1979).

In short, the vector $\underline{x}$ majorizes $\underline{y}$ if the coordinates of $\underline{x}$ are more dispersed than are the coordinates of $\underline{y}$, subject to the constraint that the sum of the coordinates of $\underline{x}$ and of $\underline{y}$ is the same.

A Schur convex function then is one that increases as dispersion increases (where the concept of dispersion used is specifically linked to the majorization order).

The extremal case under the majorization order corresponds to the choice $x_i = (\sum_{j=1}^{n} x_j)/n$. In particular then, a Schur convex function will take on a larger value when there is some variability in $\underline{x}$ than it does when there is no variability [i.e., when $x_i = \bar{x} = (\sum_{j=1}^{n} x_j)/n, i = 1, 2, \ldots, n]$.

Many examples of Schur convex functions can of course be found in the literature. Perhaps the simplest example is what is called a separable convex function. It is of the form

$$g(x) = \sum_{i=1}^{n} h(x_i),$$

where $h$ is a convex function.

We now begin our tour of examples in the literature in which majorization makes cameo and/or starring appearances.

One can even consider a variation of the children's game "Where's Waldo?". In that game a very complicated picture is provided in which, hidden away, is a picture of the hero Waldo. He is always there, but he is often hard to find. Similarly we can view various areas of statistical research and/or applications as being rather complicated scenes in which perhaps Waldo, a.k.a. majorization, may well be lurking. The search begins.

## 3. COVERING A CIRCLE WITH RANDOMLY PLACED ARCS

Suppose that $n$ arcs of lengths $\ell_1, \ell_2, \ldots, \ell_n$ are placed independently and uniformly on the unit circle (a circle with unit circumference). Let $P(\underline{\ell})$ denote the probability that the unit circle is completely covered by these arcs. The problem is only interesting when the total length of the arcs $L = \sum_{i=1}^{n} \ell_i$ exceeds 1, the circumference of the circle. We therefore assume that $L > 1$. In the special case in which the arcs are of equal lengths (say $\bar{\ell} = L/n$), the required probability was provided by Stevens (1939). Specifically we have

$$(3.1) \quad P(\bar{\ell}\underline{1}) = \sum_{k=0}^{n} (-1)^k \binom{n}{k} (1 - k\bar{\ell})_+^{n-1}.$$

At the other extreme, if one arc is of length $L$ and the others of length 0, coverage is certain. It would appear then that, in this situation, increasing the variability among the $\ell_i$'s subject to the sum being equal to $L$, might well be associated with an increase in the coverage probability. Proschan conjectured that $P(\underline{\ell})$ is a Schur convex function. It is indeed Schur convex but it is not that easy to verify. Details were provided by Huffer and Shepp (1987). Not surprisingly, the argument is based on studying the effect on $P(\underline{\ell})$ of making a small change in two unequal $\ell_i$'s (to make them more alike) holding the other lengths fixed. Waldo is here, but he is not easily unmasked.

## 4. WAITING FOR A PATTERN

If we seat a monkey at a keyboard and have him type letters, spaces and punctuation marks at random, it is common knowledge that eventually he will produce a perfectly typed version of the Gettysburg Address and, for that matter, the entire contents of the 2004 edition of the *Encyclopedia Brittanica*. But we would have to wait a rather long time to see this.

The mathematical formulation of the monkey's activities involves observing a sequence $X_1, X_2, \ldots$ of independent identically distributed random variables with possible values $1, 2, \ldots, k$ and associated positive probabilities $p_1, p_2, \ldots, p_k$. Let $N$ denote the waiting time until a particular consecutive string of outcomes is observed, or one of a particular set of outcome strings is observed. If we are waiting for the string $t_1, t_2, \ldots, t_\ell$ where each $t_j$ is a number chosen from the set $1, 2, \ldots, k$, there are several ways in which variability can affect the waiting time random variable $N$. The random variable will be affected by variability among the $p_i$'s, the probabilities of the individual possible values of the $X$'s. It will be also affected by the variability among the $t_j$'s appearing in the string whose appearance we are awaiting. For example, we might expect to have to wait longer for a string of $\ell$ consecutive like outcomes than for a string of $\ell$ distinct outcomes. Possibilities for a role for majorization abound here.

In particular, Ross (1999) considers the waiting time $N$ until we observe a run of $k$ observed values of the $X_i$'s that includes all $k$ of the possible values of the $X_i$'s, as a function of $\underline{p} = (p_1, \ldots, p_k)$. Here indeed it is possible to verify that for every $n$, $P(N > n)$ is a Schur convex function of $\underline{p}$, and consequently that $E(N)$ is also Schur convex as a function of $\underline{p}$. The shortest waiting time is thus associated with the case in which the $p_j$'s are all equal to $1/k$.

## 5. PAIRED COMPARISONS

The theory of paired comparisons has found considerable application in the study of professional sporting contests. At the end of a typical season each of the $k$ teams in the league will have played each other team a given number, say $p$, of times. For simplicity, we ignore such factors as home field advantage and we assume that the rules of the league exclude the possibility of ties. Similar analysis might well be applied to taste-testing experiments and other paired comparison scenarios, but we will follow Joe (1988) and focus on the sports setting.

In modeling this scenario, it is convenient to consider a $k \times k$ matrix $P = (p_{ij})$ in which, for $i \neq j$, $p_{ij}$ denotes the probability that team $i$ will beat team $j$ in a particular game. Of course we have $p_{ij} + p_{ji} = 1$, recalling our assumption that ties do not occur. We leave the diagonal elements of $P$ empty so that $P$ has $n(n-1)$ nonnegative elements. The strength of a particular team, say team $i$, is to some extent measured by its corresponding row total $p_i = \sum_{j \neq i} p_{ij}$. For a given vector $\underline{p}$ of team strengths, we can consider the class $\mathcal{P}(\underline{p})$ of all probability matrices $P$ with only off-diagonal elements defined and with row totals given by $\underline{p}$.

It is reasonable to assume that if team $i$ is better than team $j$ (i.e., if $p_{ij} \geq 0.5$) and if team $j$ is better than team $k$, then team $i$ should be better than team $k$.

Joe calls the matrix $P$ weakly transitive if $p_{ij} \geq 0.5$ and $p_{jk} \geq 0.5$ imply $p_{ik} \geq 0.5$. A stronger condition is also plausible. He defines $P$ to be strongly transitive if $p_{ij} \geq 0.5$ and $p_{jk} \geq 0.5$ imply $p_{ik} \geq \max(p_{ij}, p_{jk})$.

Where does majorization come into this picture? Each matrix $P$ in $\mathcal{P}(\underline{p})$ can be rearranged as an $n \times (n-1)$-dimensional row vector denoted by $P^*$. We will write $P \prec Q$ iff $P^* \prec Q^*$ in the usual sense of majorization. A matrix $P \in \mathcal{P}(\underline{p})$ is said to be minimal if $Q \prec P$ implies $Q^* = P^*$ up to rearrangement. Joe (1988) verifies that any strong transitive $P$ is minimal. Variations in which ties and home field advantage are considered are also discussed in Joe (1988).

## 6. PHASE TYPE DISTRIBUTIONS

In a continuous-time Markov chain with $(n+1)$ states, of which $n$ states $(1, 2, \ldots, n)$ are transient and state $n+1$ is absorbing, the time $T$ until absorption in state $n+1$ is said to have a phase type distribution (Neuts, 1975). Such distributions are parameterized by an initial distribution vector for the chain, $\underline{\alpha} = (\alpha_1, \alpha_2, \ldots, \alpha_n)$ (we assume that the probability of beginning in state $n+1$ is 0), and a matrix of intensities of transitions among the $n$ transient states $Q$. The elements of $Q$ satisfy $q_{ii} < 0, i = 1, 2, \ldots, n$, and $q_{ij} \geq 0, j \neq i$. In such a setting $T$ is said to have a phase type distribution with parameters $\underline{\alpha}$ and $Q$ and we write $T \sim PH(\underline{\alpha}, Q)$. A very simple example is the one in which $\underline{\alpha} = \underline{\alpha}^* = (1, 0, 0, 0, \ldots, 0)$ and $Q = Q^*$ where $q_{ii}^* = -\delta, \forall i$ and $q_{ij}^* = \delta$ for $j = i + 1$ while $q_{ij} = 0$ otherwise. In this situation the chain begins in state 1, and then successively moves through states $2, 3, \ldots, n$, spending an exponential ($\delta$) time in each state. Consequently the time to absorption, say $T^*$, will be a sum of $n$ i.i.d. exponential random variables and so $T^* \sim \text{gamma}(n, \delta)$ (in queueing contexts this is often called the Erlang distribution rather than the gamma distribution).

We say that a phase type distribution is of order $n$ if $n$ is the smallest integer such that the distribution can be identified with the absorption time of a chain with $n$ transient states and one absorbing state. It appears that, in some sense, $T^*$ exhibits the most regular behavior of any phase type distribution of order $n$. This can be made precise in terms of what is called the Lorenz order, a natural extension of majorization.

Let $\mathcal{L}$ denote the class of nonnegative random variables with finite positive expectations. (This can be extended to allow the random variables to assume negative values, but for our present purposes this is not needed.) For $X$ and $Y$ in $\mathcal{L}$, we will write $X \leq_L Y$ iff $E(g(X/E(X))) \leq E(g(Y/E(Y)))$ for every continuous convex function $g$. Majorization can be identified as a special case here by choosing $X$ and $Y$ to each have $n$ equally likely values $x_1, x_2, \ldots, x_n$ and $y_1, y_2, \ldots, y_n$, respectively, with $E(X) = E(Y)$. More detailed discussion of the Lorenz order on $\mathcal{L}$ may be found in Arnold (1987). Aldous and Shepp (1987) showed that $T^*$ [with its gamma$(n, \delta)$ distribution] has the smallest coefficient of variation among phase type distribution of order $n$, that is, it minimizes $E((\frac{T}{E(T)})^2)$. More generally, O'Cinneide (1991) verified that $T^* \leq_L T$ for any variable $T$ that is phase type of order $n$, thus confirming the fact that $T^*$ exhibits the least "variability" (as measured by the Lorenz order).

## 7. CATCHABILITY

An island community contains an unknown number $\nu$ of species of butterflies. Butterflies are sequentially trapped until $n$ individuals have been captured. Denote by $r$, the number of distinct species represented among the captured butterflies. We may well use $r$ (and $n$) to help us estimate $\nu$.

A typical stochastic model for this problem is based on the assumption that butterflies from species $j, j = 1, 2, \ldots, \nu$, enter the trap according to a Poisson $(\lambda_j)$ process and that these Poisson processes are independent. Define $p_j = \lambda_j / \sum_{i=1}^{\nu} \lambda_i$. The probability that a particular butterfly trapped is from species $j$ is then given by $p_j, j = 1, 2, \ldots, \nu$. The $p_j$'s can be interpreted as measures of "catchability" of the various species. The simplest model is that of equal catchability (i.e., $p_j = 1/\nu, j = 1, 2, \ldots, \nu$). If we assume that $\nu \leq n$, then, under the equal catchability model, a minimum variance unbiased estimate of $\nu$, based on $r$, exists. It is given by

(7.1) $$\hat{\nu} = S(n+1, r)/S(n, r)$$

where $S(n, x)$ is a Stirling number of the second kind. What happens when the species vary in catchability? In an extreme case in which one particular species is easily trapped and the others are extremely difficult to trap, we will usually observe $r = 1$ and will consequently badly underestimate $\nu$. Indeed as Nayak and Christman (1992) observe, the random number $R$ of species captured has a distribution which is a Schur convex function of $\underline{p}$. Thus the estimate (7.1) and other estimates which are sensible under equal catchability will be negatively biased with the bias increasing as the catchability becomes more variable.

## 8. DISEASE TRANSMISSION

Tong (1997) identifies an interesting majorization feature of a disease transmission model due to Eisenberg (1991). Consider a closed population of $n + 1$ individuals. One individual (number $n + 1$) is susceptible to the disease but as yet is uninfected. The other $n$ individuals are carriers of the disease. If individual $n + 1$ has a single contact with individual $i$, we denote the probability of avoiding infection by $p_i, i = 1, 2, \ldots, n$.

It is assumed that individual $n + 1$ makes a total of $J$ contacts with individuals in the population in accordance with a preference vector $\underline{\alpha} = (\alpha_1, \alpha_1, \alpha_2, \ldots, \alpha_n)$, where $\alpha_i > 0, i = 1, 2, \ldots, n$, and $\sum_{i=1}^{n} \alpha_i = 1$. In addition, individual $n + 1$ has a lifestyle vector $\underline{k} = (k_1, k_2, \ldots, k_J)$ where the $k_i$'s are nonnegative integers summing to $J$. For given vectors $\underline{\alpha}$ and $\underline{k}$, the individual $n + 1$ proceeds as follows. He/she first picks a partner from among the $n$ carriers according to the preference vector $\underline{\alpha}$. Thus he/she will select individual 1 with probability $\alpha_1$, individual 2 with probability $\alpha_2$, and so on. He/she then makes $k_1$ contacts with this partner. Then he/she selects a second partner (it could be the same one) according to the preference vector $\underline{\alpha}$ and has $k_2$ contacts with this partner. The process terminates after all $J = \sum_{i=1}^{J} k_i$ contacts have been made. Denote the probability of escaping infection by $H(\underline{k}, \underline{\alpha}, \underline{p})$, depending as it does on lifestyle ($\underline{k}$), preference ($\underline{\alpha}$) and variable nontransmission probabilities ($\underline{p}$).

There are several possible roles for majorization here. Variability among the coordinates of $\underline{k}, \underline{\alpha}$ and/or $\underline{p}$ can be expected to affect $H(\underline{k}, \underline{\alpha}, \underline{p})$. Tong (1997) focuses on the lifestyle vector $\underline{k}$. Two extreme lifestyles are readily identified. The first one corresponds to $\underline{k} = (J, 0, 0, \ldots, 0)$ which could be called a monogamous style. Here a partner is randomly chosen according to the preference vector $\underline{\alpha}$ and all contacts are made with this individual. The second extreme lifestyle has $\underline{k} = (1, 1, 1, \ldots, 1)$. In this case each contact is made with a randomly chosen individual. The probability of escaping infection with $\underline{k} = (J, 0, \ldots, 0)$ is clearly $\sum_{i=1}^{n} \alpha_i p_i^J$ while the probability of escaping infection using the lifestyle $(1, 1, 1, \ldots, 1)$ is $(\sum_{i=1}^{n} \alpha_i p_i)^J$. It follows via Jensen's inequality that one has a larger probability of escaping infection with the "monogamous" lifestyle $(J, 0, \ldots, 0)$ than with the "random" lifestyle $(1, 1, 1, \ldots, 1)$. This holds for every $\underline{\alpha}$ and every $\underline{p}$. But of course these two lifestyles are extreme cases with regard to majorization. It is then quite plausible that the probability of escaping infection is a Schur convex function of the lifestyle vector $\underline{k}$. Indeed, Tong (1997) confirms this conjecture. He also is able to get some results when the number $J$ of contacts is a random variable. Several interesting aspects of this problem remain open.

## 9. APPORTIONMENT IN PROPORTIONAL REPRESENTATION

The ideal of one man–one vote is often approached by the device of proportional representation. Thus if there are $N$ seats available and if a political party received $100q\%$ of the votes, then ideally that party should be assigned $Nq$ seats. But fractional seats cannot be assigned (or better yet *are* not assigned, since there seems to be no reason why they could not be assigned, except perhaps for aesthetic considerations). Which method of rounding should be used to arrive at an assignment of integer-valued numbers of seats to every party in a manner essentially reflecting proportional representation? This is not a new problem. Several very well-known American politicians have proposed methods of rounding for use in this situation. Balinski and Young (2001) provide a good survey of the methods usually considered. Marshall, Olkin and Pukelsheim (2002) highlight the role of majorization in comparing the various candidate rounding methods. John Quincy Adams proposed a method that was kind to small parties (rounding up their representation), while at the other extreme Thomas Jefferson urged rounding down, which favors large parties. Other popular intermediate strategies are associated with the names Dean, Hill and Webster.

It is easiest to describe all of these apportionment methods in terms of a sequence of signposts which determine rounding decisions. The signposts $s(k)$ are numbers in the interval $[k, k+1]$ such that $s(k)$ is a strictly increasing function of $k$. The corresponding rounding rule is that a number in the interval $[k, k+1]$ is rounded down if it is less than $s(k)$ and is rounded up if it is greater than $s(k)$. If the number is exactly equal to $s(k)$, then we may round up or down. So-called power-mean signpost sequences have been popular. They are of the form

$$(9.1) \quad s_p(k) = \left( \frac{k^p}{2} + \frac{(k+1)^p}{2} \right)^{1/p},$$
$$-\infty \leq p \leq \infty.$$

The five most popular apportionment methods can all be viewed as having been based on a particular power-mean signpost sequence. The Adams rule (rounding up) corresponds to $p = -\infty$, the Dean rule corresponds to $p = -1$, the Hill rule corresponds to $p = 0$, the Webster rule to $p = 1$ and finally the Jefferson rule (rounding down) corresponds to $p = \infty$. Marshall, Olkin and Pukelsheim (2002) show that the seating vector produced by a power-mean rounding rule of order $p$ will always be majorized by the seating vector produced by a power-mean rounding rule of order $p'$ if and only if $p \leq p'$. Consequently, among the five popular apportionment rules, the change when moving from the Adams rule toward the Jefferson rule is a change in favor of large parties in a majorization sense. The move from an Adams apportionment toward a Jefferson apportionment can actually be accomplished by a series of single seat reassignments from a poorer party (with fewer votes) to a richer party (with more votes) [paralleling reverse Robin Hood (a.k.a. Pigou–Dalton) income transfers in an economic setting].

## 10. MAJORIZATION IN STATISTICAL MECHANICS

The state space of a physical system, $S_n$, can be identified with the set of all probability vectors $\underline{p} = (p_1, p_2, \ldots, p_n)'$ where $p_i \geq 0$ and $\sum_{i=1}^n p_i = 1$. A useful partial order in this context is related to the information content of the states. For two states $\underline{p}$ and $\underline{q}$, it is prescribed that $\underline{p} \prec \underline{q}$ iff there exists a doubly stochastic matrix $T$ with $\underline{p} = T\underline{q}$. But of course, appealing to the classical result of Hardy, Littlewood and Pólya (1929), this is in fact the majorization partial order (and the notation is thus consistent with our usage in earlier sections of this paper). In this context separable concave functions are called generalized entropies.

A related partial order is defined on $k$-tuples of states. For two $k$-tuples $(\underline{p}_1, \underline{p}_2, \ldots, \underline{p}_k)$ and $(\underline{q}_1, \underline{q}_2, \ldots, \underline{q}_k)$ we define

$$(\underline{p}_1, \underline{p}_2, \ldots, \underline{p}_k) \prec^{(k)} \underline{q}_1, \underline{q}_2, \ldots, \underline{q}_k)$$

iff there exists a stochastic matrix $T$ such that $\underline{p}_i = T\underline{q}_i, i = 1, 2, \ldots, k$. In particular when $k = 2$, a partial ordering defined with respect to a reference state $\underline{s}$ becomes of interest. The partial order relative to $\underline{s}$ is defined by

(10.1) $\quad \underline{p} \prec_{\underline{s}} \underline{q} \quad \text{iff } (\underline{p}, \underline{s}) \prec^{(2)} (\underline{q}, \underline{s}).$

It may be noted that if $\underline{s}$ is chosen to be equal to $\underline{e} = (\frac{1}{n}, \ldots, \frac{1}{n})$, then the corresponding partial order (relative to $\underline{e}$) coincides with the usual majorization order. Thus the partial ordering $\prec_{\underline{s}}$ is a genuine extension of the classical majorization order.

Dynamic processes in the state space $S_n$ can be identified with indexed families of stochastic matrices. Such processes which preserve the $\underline{s}$-partial order have been studied in some detail. A convenient introductory reference is Zylka (1985).

Schur convex functions and analogous $\underline{s}$-Schur convex functions turn out to have useful thermodynamic interpretation in this context.

## 11. CONNECTED COMPONENTS IN A RANDOM GRAPH

Ross (1981) considers a random graph with nodes numbered $1, 2, \ldots, n$. Suppose that $X(1), X(2), \ldots, X(n)$ are independent identically distributed random variables each with possible values $1, 2, \ldots, n$ and with common distribution defined by

(11.1) $\quad P(X(i) = j) = p_j, \quad j = 1, 2, \ldots, n,$

where $p_j \geq 0, \forall j$ and $\sum_{j=1}^n p_j = 1$. We construct the random graph by drawing the $n$ random arcs $(i, X(i))$, $i = 1, 2, \ldots, n$. In this manner, one arc emanates from each node. However, of course, several arcs can terminate at the same node. The resulting graph will have a random number of connected components. A connected component of the graph is a set of nodes such that any pair of them is linked by an arc in the graph, and there are no arcs joining any nodes in the set with any node outside the set. Let us denote the random number of such connected subsets by $M$. The distribution of $M$ will of course be influenced by the probability vector $\underline{p}$, appearing in (11.1), which governs the distribution of the random arcs $X(1), X(2), \ldots, X(n)$.

For example, if $\underline{p} = (1, 0, 0, \ldots, 0)$, then all arcs will terminate at node 1 and there will be a single connected subset of nodes in the random graph, that is, $M = 1$.

The following expression for the expected value of $M$ is provided by Ross:

(11.2) $\quad E(M) = \sum_S (|S| - 1)! \prod_{j \in S} p_j$

where the summation extends over all nonempty subsets of $\{1, 2, \ldots, n\}$. It is then possible, using this expression, to verify that $E(M)$ is a Schur concave function of $\underline{p}$. Consequently the expected number of connected components of the graph is maximized if $p_j = 1/n, j = 1, 2, \ldots, n$.

## 12. A STOCHASTIC RELATION BETWEEN THE SUM OF TWO RANDOM VARIABLES AND THEIR MAXIMUM

Suppose that $\underline{X} = (X_1, X_2)$ is a random vector with nonnegative coordinate random variables $X_1, X_2$. It is often of interest to compare the tail behavior of $X_1, X_2$ with that of $\max(X_1, X_2)$. In the context of construction of confidence intervals for the difference between normal means with unequal variances (a Behrens–Fisher setting), Dalal and Fortini (1982) identified a sufficient condition for stochastic ordering between $X_1 + X_2$ and $\sqrt{2}\max(X_1, X_2)$ that involves Schur convexity. Specifically they prove that a sufficient condition for

$$P(X_1 + X_2 \leq c) \geq P(\sqrt{2}\max(X_1, X_2) \leq c)$$

for any $c \geq 0$, is that the joint density of $(X_1, X_2)$, say $f(x_1, x_2)$, is such that $f(\sqrt{x_1}, \sqrt{x_2})$ is a Schur convex function of $\underline{x}$. The proof involves conditioning on $X_1^2 + X_2^2$ and observing that on any curve $x_1^2 + x_2^2 = t$, the joint density $f(x_1, x_2)$ increases as one moves away from the line $x_1 = x_2$.

An important special case in which the hypotheses are satisfied is the situation in which $(X_1, X_2) = (|Y_1|, |Y_2|)$ where $\underline{Y} \sim N^{(2)}(\underline{0}, \sigma^2 \begin{pmatrix} 1 & \rho \\ \rho & 1 \end{pmatrix})$.

A related $n$-dimensional result is also provided by Dalal and Fortini (1982). They show that if $X_1, X_2, \ldots, X_n$ are i.i.d. positive random variables with common density $f$ and if $\log f(\sqrt{x})$ is concave and $f(x)/x$ is nonincreasing, then

$$\sum_{i=1}^{n} X_i \leq_{st} \sqrt{n} \max(X_1, X_2, \ldots, X_n).$$

## 13. FURTHER EXAMPLES

The list could be continued. Schur convexity and majorization can be found in many other settings. To conclude our short survey, we will merely mention briefly a few more interesting settings in which Waldo appears:

(i) the study of peakedness of univariate and multivariate distributions,

(ii) admissibility of tests in multivariate analysis of variance,

(iii) probability content of regions for a Schur concave joint density,

(iv) the study of diversity in ecological environments,

(v) income and wealth inequality measurement (with multivariate extensions).

As observed in the Introduction, there are many more examples in the literature and there is no reason to believe that the search for new applications of majorization and Schur convexity will falter in the next 25 years. When the *Inequalities* volume celebrates its golden jubilee, an even more extensive and fascinating array of appearances can be confidently predicted. The search for Waldo will continue apace.